\documentclass[11pt,a4paper]{article}

\usepackage[T1]{fontenc}
\usepackage[utf8]{inputenc}
\usepackage{lmodern}
\usepackage[a4paper,margin=1in]{geometry}
\usepackage{microtype}
\usepackage{amsmath,amssymb}
\usepackage{graphicx}
\usepackage{booktabs}
\usepackage{siunitx}
\usepackage{caption}
\usepackage{subcaption}
\usepackage{float}
\usepackage{url}
\usepackage[hidelinks]{hyperref}
\usepackage{authblk}

\graphicspath{{figures/}}
\title{Electrodeposited Co and Ni Hexacyanoferrates: Insights into Structure and Morphology}

\author[1]{Larissa de O. Garcia}
\author[1]{Michael Pohlitz}
\author[2]{Mohammed F. Kalady}
\author[2]{Falk R\"oder}
\author[2,3]{Axel Lubk}
\author[2]{Daniel Wolf}
\author[1]{Christian K. M\"uller\thanks{Corresponding author: Christian.Mueller.1@whz.de}}

\affil[1]{Faculty of Physical Engineering/Computer Sciences, University of Applied Sciences Zwickau, 08056 Zwickau, Germany}
\affil[2]{Leibniz Institute for Solid State and Materials Research Dresden, 01069 Dresden, Germany}
\affil[3]{Institute of Solid State and Materials Physics, TU Dresden, Haeckelstra\ss e 3, 01069 Dresden, Germany}

\date{}

\begin{document}
\maketitle

\begin{abstract}
Prussian blue (PB) and its analogues (PBAs) are interesting materials for electrochemical applications due to their tunable redox chemistry and open framework structure. In this study, hexacyanoferrates (HCF) containing iron (FeHCF), cobalt (CoHCF) and nickel (NiHCF) were synthesized via potentiostatic electrodeposition. Cyclic voltammetry revealed distinct redox behaviors. Morphological characterization (SEM, EDX) demonstrated uniform, pyramidal film growth for FeHCF and CoHCF. Otherwise, NiHCF presented a cracked film with cubic clusters on top due to residual stress. Despite this, homogeneous element distribution was found for all samples. Structural characterization (TEM and XRD) confirmed a cubic lattice crystal structure for all films, with systematic lattice contraction from Fe to Co to Ni due to decreasing atomic radius. Raman and XPS data revealed a shift toward Fe$^{2+}$ dominant oxidation states and modifications in C$\equiv$N bonding with the influence of K$^+$ and water occupancy in the PBAs framework. These findings illustrate how metal substitution and deposition parameters can tune the structural and electrochemical properties of PBA films, presenting a strategic route to design tailored electrodes.
\end{abstract}


\section{Introduction}
Since the first report of Prussian Blue (PB) layer growth using a solid electrode by V. D. Neff in 1978 \cite{r1}, PB films have been extensively studied and applied in electronic and energy fields such as supercapacitors \cite{r2}, electrocatalysis \cite{r3,r4}, biosensors \cite{r5}, electrochromic devices \cite{r6,r7}, hydrogen storage \cite{r8} and batteries due to their facile and tunable physicochemical properties \cite{r9}.

With a general formula described as A$_x$M$_A$[M$_B$(CN)$_6$]$_y$nH$_2$O, Prussian blue materials correspond to the hexacyanoferrate class, where A is an alkali metal cation (e.g. K$^+$, Na$^+$, Li$^+$, NH$_4^+$) and M$_A$/M$_B$ denote transition metal elements connected with the cyano groups present in the structure (M$_A$ = Mn, Fe, Co, Ni, Cu, Zn; M$_B$ = Fe, Co, Cr). The crystal structure of PB was first described by Keggin and Miles \cite{r10} as a cubic framework composed of Fe$^{2+}$--C$\equiv$N--Fe$^{3+}$ chains, with six cyano groups linked to each Fe ion octahedrally coordinated by six cyanide ligands, resulting in a cubic lattice with a parameter of approximately 10.2~\AA \cite{r11,r12,r13}. In cases where Fe is partially substituted by another transition metal, so-called Prussian blue analogues (PBAs) are formed. Such substitutions can induce distortions in the cubic structure, leading to lattice parameters of around 10--10.5~\AA, depending on the ionic radius of the M$_B$ metal when a Fe atom is substituted by a larger transition metal \cite{r10,r14}.

Numerous synthetic methods have been described over the years for producing PB and PBA materials such as co-precipitation \cite{r15,r16}, hydrothermal synthesis \cite{r15}, ion exchange \cite{r16}, microemulsion \cite{r17} and electrodeposition \cite{r18}. Among these methods, electrodeposition offers advantages including direct growth on conductive substrates, control over film formation and compatibility with electrochemical tuning of the deposited layers \cite{r17,r18,r19,r20}. Following the initial report in 1978 \cite{r1}, in which a PB film was deposited on a gold substrate, Itaya et al. \cite{r21} introduced a galvanostatic procedure that enabled the fabrication of high-quality PB films on various electrode substrates using acidic solutions FeCl$_3$ and K$_3$Fe(CN)$_6$ as precursor \cite{r6,r19,r20,r22,r23}. Since then, several research groups have investigated electrodeposited PB and related materials for electronic applications, including resistive-switching and memristive devices \cite{r19,r20,r24,r25,r26,r27,r28,r66,r67}. However, only a limited number of studies have addressed the electrodeposition of PBAs. These contributions provide important insights into their electrochemical behavior but often lack detailed structural and morphological characterization \cite{r29,r30,r31,r32}. For this reason, this work presents a comprehensive structural, morphological and compositional characterization of different PBA layers, including cobalt- and nickel-based films electrodeposited on gold substrates, with the aim of elucidating the impact of these metals on the Prussian blue crystal structure. The findings reported here might provide insights for the future development and application of Prussian blue analogues in electronic devices.

\section{Materials and Methods}
\subsection{Sample Preparation}
Prussian Blue Analogue films were grown until total deposited charge of 30~mC was reached by applying a constant potential of 0.30~V using an Ivium electrochemical workstation (Ivium CompactStat, Eindhoven, The Netherlands). The potentiostatic fabrication of these Hexacyanoferrate layers (HCF) was carried out with a three-electrode setup (Pt as counter electrode, saturated calomel electrode (SCE) as reference electrode and 50~nm Au/5~nm Cr on Si (100) as working electrode) at 25~$^\circ$C. All potentials reported for the electrochemical deposition and cyclic voltammetry experiments are referenced to the SCE.

The deposition of the layers occurred in a circular area of $\sim$0.5~cm$^2$ defined by a mask of adhesive tape on the surface of the working electrode and aqueous solutions containing 1.0~M KCl (ACS, 99--100.5\%, Sigma Aldrich, Darmstadt, Germany), 0.25~mM K$_3$Fe(CN)$_6$ (ACS >99\%, Sigma Aldrich, Darmstadt, Germany), and 0.25~mM metal chloride salts were used as electrolyte for PBA deposition. The electrolyte solution pH was set around 2 using HCl (ACS, 37\%, Sigma Aldrich, Darmstadt, Germany) for all deposition. These procedures were previously reported for electrodeposited Prussian Blue-based films \cite{r19,r33,r67}.

In this work samples were named according to the metal chloride salt used in electrolyte solution, such as FeHCF for iron chloride hexahydrate (FeCl$_3\cdot$6H$_2$O, ACS, 98--102\%, Sigma Aldrich, Darmstadt, Germany), CoHCF for cobalt chloride hexahydrate (CoCl$_2\cdot$6H$_2$O, ACS, 98.0--102.0\%, Thermo Scientific Chemicals, Germany) and NiHCF for nickel chloride hexahydrate (NiCl$_2\cdot$6H$_2$O, $\geq$99.9\%, Sigma-Aldrich, Germany).

\subsection{Material Characterization}
Electrodeposited HCF layers were characterized by X-ray diffraction performed with a X'Pert MRD diffractometer (Panalytical, The Netherlands) using Co-K$\alpha$ radiation with a wavelength of $\lambda=1.78896$~\AA in Bragg--Brentano geometry with a step-size of $0.005^\circ$ in the range of $2\theta=10$--$50^\circ$. Morphological and compositional properties were analyzed by field emission scanning electron microscopy (FEG-SEM, TESCAN CLARA, Brno, Czech Republic) equipped with an energy-dispersive X-ray (EDX) detector (Ultim Max 65 SDD, Oxford Instruments, Wiesbaden, Germany) at 20~keV and a Raman system (Witec RISE, Ulm, Germany). Raman measurements were performed with a 532~nm laser at 0.4~mW.

In order to conduct cross-sectional transmission electron microscopy (TEM), ca. 50~nm thick lamellae were prepared using focused ion beam (FIB) milling. High-resolution TEM (HRTEM) was conducted on a double-corrected FEI Titan$^3$80-300 TEM instrument (Thermo Fisher Scientific, US) operated at 300~keV to investigate the crystal structure of the films at the atomic scale. In bright-field TEM (BFTEM) mode, an objective aperture of 50~$\mu$m diameter was used. Scanning transmission electron microscopy (STEM) and electron energy loss spectroscopy (EELS) were carried out using a Hitachi HF3300 S TEM operated at 300~keV, equipped with a CEFID image filter \cite{r34} including a TVIPS XF416 detector. The energy dispersion was selected to realize an energy range of 1024~eV and resolution of 1.3~eV. Elemental maps were extracted for Fe, Co, Ni, K, C, N, and O using the background fitting and core-loss quantification procedures integrated in Gatan Inc. DigitalMicrograph image analysis software. X-ray photoelectron spectroscopy (XPS, SPECS, Berlin, Germany) was performed using an Al K$\alpha$ radiation source (photon energy = 1486.7~eV). The size of the investigated area of interest from the surface was $\sim$1~mm and before analysis the samples were sputtered with argon ions at 3~keV at a pressure of $5.10\times10^{-8}$~mbar to remove surface contaminants. The removed layer thickness is estimated to be around 10--20~nm and the spectral analysis was performed with Casa (CasaXPS, vs. 2.2.24, USA).

\section{Results and Discussion}
\begin{figure}[htbp]
  \centering
  \includegraphics[width=0.70\linewidth]{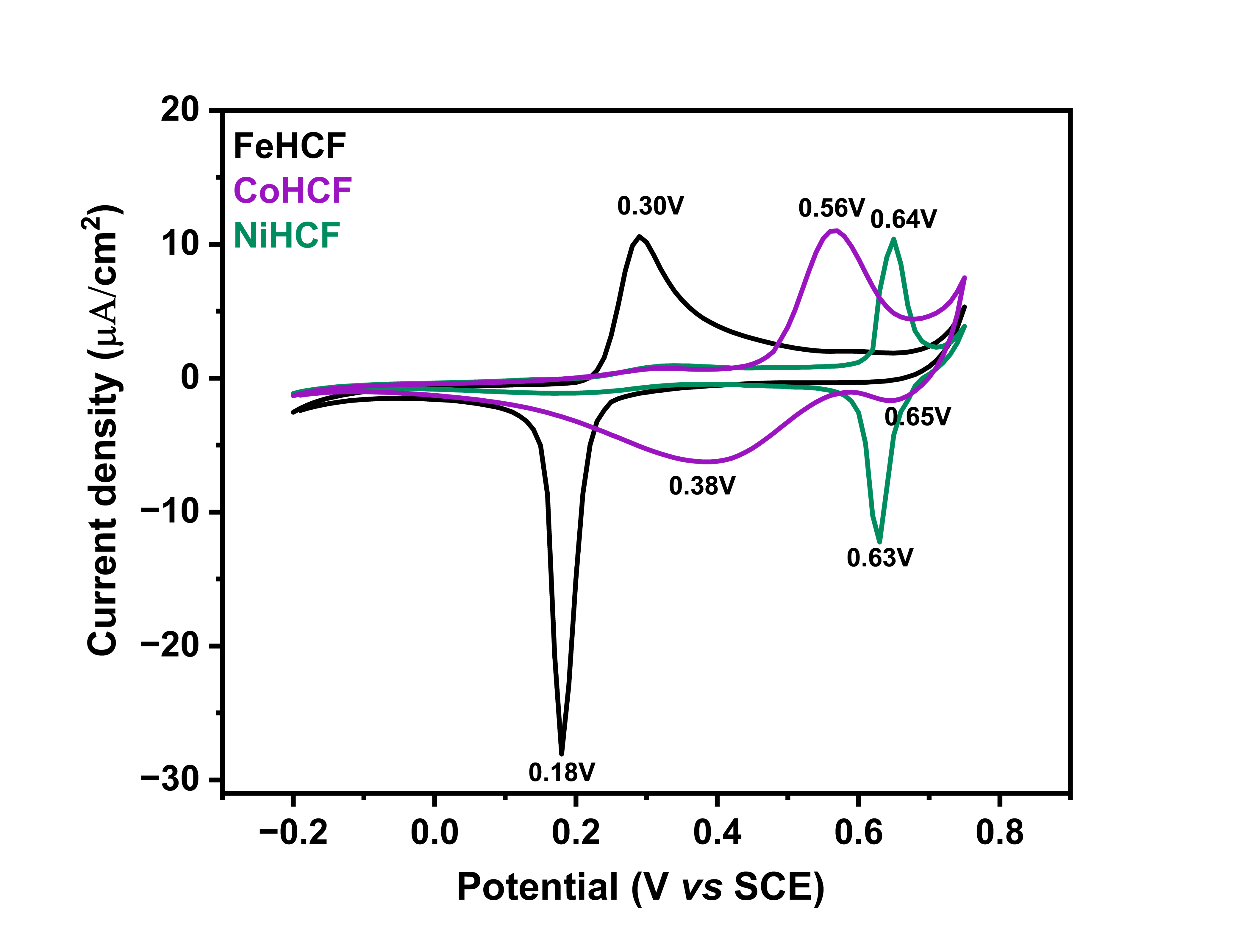}
  \caption{Cyclic voltammetry curves obtained at a scan rate of 100~mV/s for FeHCF, CoHCF and NiHCF.}
  \label{fig:cv}
\end{figure}

It is well established that the chosen potential for deposition will affect the quality and oxidation state of the Prussian blue films \cite{r25,r67}. The optimal electrodeposition conditions for PB analogues containing cobalt and nickel were determined by means of cyclic voltammetry (CV) in the potential window ranging from $-0.2$~mV to 0.6~V at a scan rate of 100~mV/s. The CV obtained for electrolytes containing different transition metals is shown in Figure~\ref{fig:cv}.

FeHCF exhibits the characteristic Fe$^{3+}$/Fe$^{2+}$ redox behavior associated with PB formation at approximately 0.30~V (oxidation) and 0.18~V (reduction) \cite{r33}. In contrast, CoHCF displays more complex electrochemical behavior, with one oxidation peak at 0.56~V and two reduction peaks at 0.38~V and 0.65~V, suggesting the presence of two electroactive forms. NiHCF, on the other hand, presents a single oxidation and reduction process around 0.64~V and 0.63~V. These processes are attributed to the Fe$^{3+}$/Fe$^{2+}$ redox site and are consistent with the negative standard potential of Ni$^{2+}$ \cite{r35,r36}. Notably, the Fe$^{3+}$/Fe$^{2+}$ redox couple for Co and Ni analogues is shifted with respect to FeHCF, which can be attributed to the influence of K$^+$ ions present in the electrolyte solution. Cations such as K$^+$ can modulate local electronic environments and shift redox potentials as previously demonstrated in related systems by Phadke et al. \cite{r37}. Recent nanoscale studies of electrodeposited PB analogues further linked K$^+$ redistribution to Fe$^{2+}$/Fe$^{3+}$ redox reconfiguration and electrical conductance \cite{r67}. Based on these CV observations, a 0.30~V deposition potential was selected for all films to ensure consistency and comparability.

Given the variations in redox and crystallization mechanisms observed among Co- and Ni-containing films, we further conducted morphological and structural characterizations to elucidate these effects.

\begin{figure}[htbp]
  \centering
  \includegraphics[width=0.90\linewidth]{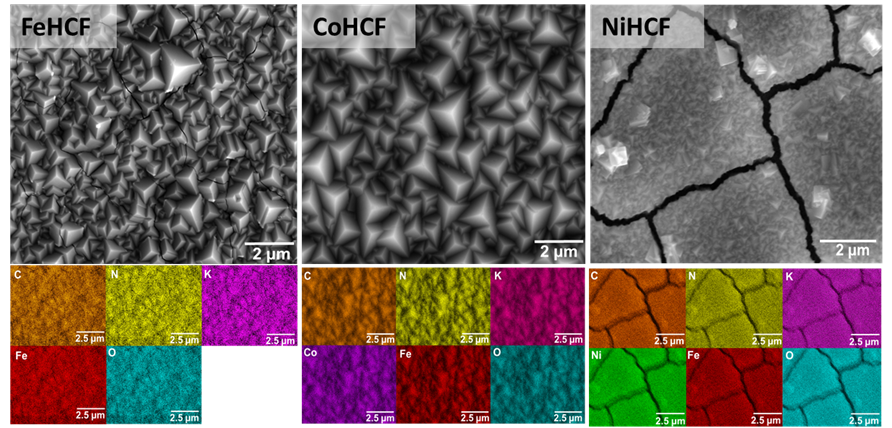}
  \caption{SEM surface image from sample top view with elemental mapping composition for FeHCF, CoHCF, NiHCF.}
  \label{fig:sem}
\end{figure}

Figure~\ref{fig:sem} shows SEM top view sections of the layers with elemental distribution for FeHCF, CoHCF, NiHCF. In terms of crystallization, FeHCF and CoHCF samples present the characteristic pyramidal growth with uniform coverage as previously described in literature \cite{r31,r36}. Cracks in FeHCF appear under electron beam due to the rapid volatilization of organic cations and water molecules on the film surface \cite{r38}.

In contrast, NiHCF sample shows non-uniform growth with formation of clusters on the surface. A previous study published by Malik et al. \cite{r39} indicate that the efficiency of NiHCF thin film growth is higher only in the first layer due to the more effective reduction of Fe(CN)$_6$ at the gold surface than at the electrode consisting of gold and a single layer of NiHCF. Decreases in reduction rate and changes in mass transfer during the electrochemical process implies the formation of clusters on the surface and increased residual stress causing long term instability and cracks after film deposition and during SEM imaging \cite{r40}.

\begin{figure}[htbp]
  \centering
  \includegraphics[width=0.90\linewidth]{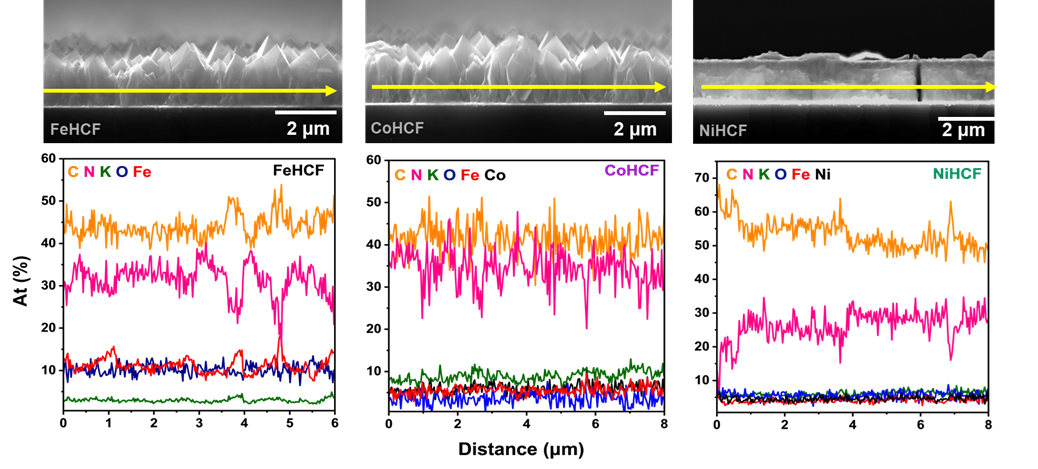}
  \caption{Cross-section SEM images of the different HCF complexes with line profiles of the elemental distribution in atomic percentage (At(\%)) obtained from EDX spectra. The yellow arrows in the SEM images indicate the line scan positions.}
  \label{fig:edxline}
\end{figure}

Despite these morphological variations, EDX analysis reveals a homogeneous distribution of constituent elements in all three samples, as the atomic compositions shown in Table~\ref{tab:edx} support. Notably, the CoHCF and NiHCF samples exhibit a significant increase in potassium content compared to FeHCF, which suggests that substitution of the transition metal may favor higher K$^+$ incorporation into the structure. Recent work on electrodeposited PB analogues has directly associated K$^+$ redistribution with redox-state changes and conductance modulation \cite{r67}. Cross-sectional SEM with EDX line scans (Figure~\ref{fig:edxline}) confirm a consistent elemental composition throughout the film. Abrupt variations in the line profiles can be attributed to cracks (NiHCF) and grain boundaries.

The higher potassium content in CoHCF and NiHCF suggests a more pronounced presence of interstitial K$^+$ ions, potentially leading to lattice distortion. The ratio between metallic ions and potassium indicates changes in the oxidation state of the samples, as well as the presence of a mixture of Prussian Blue phase with its reduced form Prussian White \cite{r41,r42}. In addition to possible changes in oxidation state due an increase in potassium amount and the presence of interstitial water indicated by oxygen traces, the metallic substitution observed in the EDX analysis could cause distortions in crystal lattice \cite{r43,r44,r45,r46}.

\begin{table}[htbp]
\centering
\caption{Elemental mapping quantification (atomic \%) of FeHCF, CoHCF and NiHCF thin films obtained by EDX.}
\label{tab:edx}
\begin{tabular}{lrrrrrrr}
\toprule
Material & C (\%) & N (\%) & O (\%) & Fe (\%) & Co (\%) & Ni (\%) & K (\%)\\
\midrule
FeHCF & 41.3 & 35.8 & 6.4 & 12.6 & --- & --- & 3.8\\
CoHCF & 38.0 & 34.2 & 5.6 & 5.6 & 6.0 & --- & 10.3\\
NiHCF & 43.7 & 38.1 & 1.8 & 4.4 & --- & 5.0 & 7.0\\
\bottomrule
\end{tabular}
\end{table}

To verify these possible changes of the crystal lattice structure, TEM in bright-field (BFTEM) and high-resolution mode (HRTEM) as well as X-ray diffraction (XRD) were carried out. The main observations of FeHCF, CoHCF and NiHCF thin films are summarized in Figure~\ref{fig:hrtem}.

\setcounter{figure}{4}
\begin{figure}[htbp]
  \centering
  \includegraphics[width=0.80\linewidth]{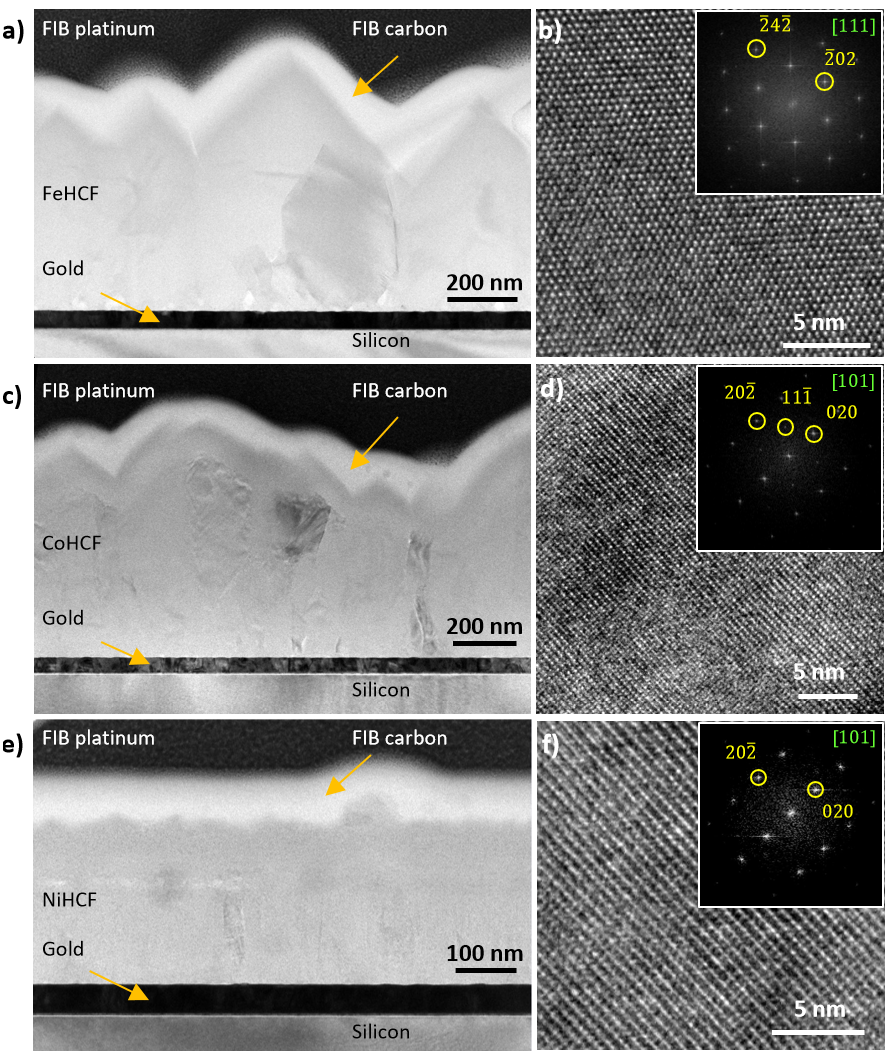}
  \caption{Cross-sectional TEM of FeHCF, CoHCF and NiHCF thin films. (a,c,e) Overview BFTEM images of the FIB-prepared lamellas of FeHCF, CoHCF and NiHCF. The dark areas show grains oriented along a certain low-index zone axis. (b,d,f) HRTEM images of the three samples showing lattice fringes with their Fourier transforms in the insets, oriented along the [111], [101] and [101] zone axes, respectively, confirming their cubic structure.}
  \label{fig:hrtem}
\end{figure}

Figure~\ref{fig:hrtem} presents the BFTEM image of a cross-section through the FeHCF layer. The poly-crystalline film appears continuous, with no visible voids or delamination at the Au interface and has an average thickness of 700~nm. Regions of increased contrast within the film indicate local areas where FeHCF grains are oriented along specific zone axes. The HRTEM image, acquired along [111] zone axis, reveals uniform and well-resolved lattice fringes within the field of view, confirming the high crystallinity of the film. The corresponding fast Fourier transform (FFT) pattern displays distinct reflections indexed to the $(\bar{2}02)$ and $(\bar{2}4\bar{2})$ planes.

The BFTEM image of the CoHCF cross-section shows a poly-crystalline film with a mean thickness of approximately 1~$\mu$m. The HRTEM image, acquired along the [101] zone axis, exhibits uniform and well-defined lattice fringes within the field of view. The corresponding FFT pattern shows reflections indexed to the $(\bar{1}11)$, (020), and $(20\bar{2})$ planes.

For NiHCF, the BFTEM image also reveals a continuous poly-crystalline film with an estimated mean thickness of 300~nm. The HRTEM image, recorded along the [101] zone axis, exhibits lattice fringes that correspond in the FFT pattern to reflections indexed to the (020) and $(20\bar{2})$ planes.

Analysis of multiple TEM images of all the three films (FeHCF, CoHCF, NiHCF) yielded lattice constants of 10.0--10.5~\AA. To this end, first, the reflection distances to the zero beam in the FFTs of the TEM images, which correspond to the reciprocals of the lattice fringe spacings $d_{hkl}$, were measured. Then, the relation
\begin{equation}
 d_{hkl}=\frac{a}{\sqrt{h^2+k^2+l^2}}
\end{equation}
where $h$, $k$, $l$ represent the Miller indices was used to determine the lattice parameter $a$ of the cubic crystals. In fact, TEM measurements lacked the precision required to identify significant changes in lattice constant attributable to ionic radius differences.

\begin{figure}[htbp]
  \centering
  \includegraphics[width=0.80\linewidth]{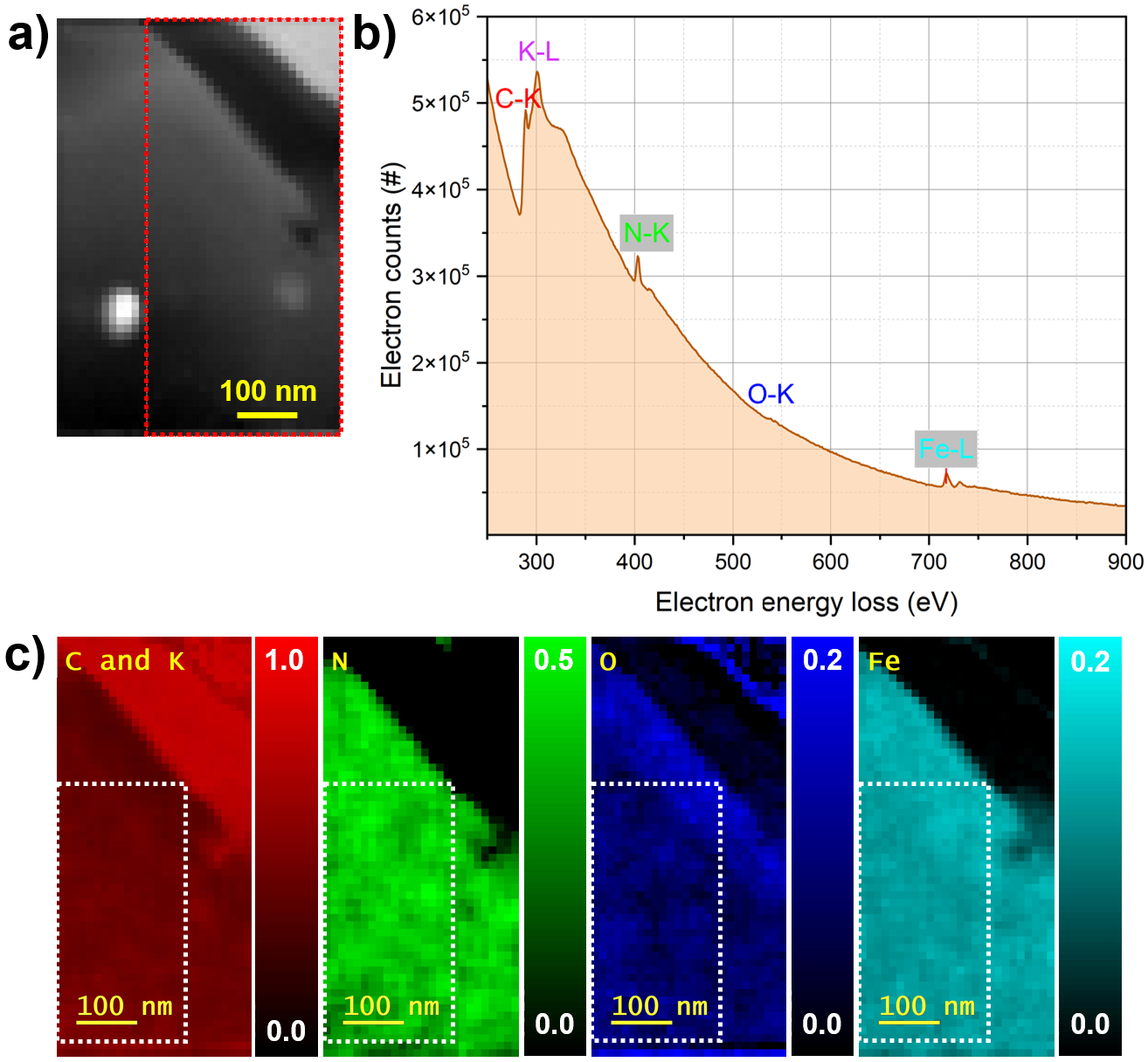}
  \caption{EELS analysis of FeHCF. (a) HAADF-STEM image of the FeHCF lamella with red box indicating the region used for EELS analysis. (b) EEL spectrum showing distinct edges for C, K, N, O, and Fe. (c) Elemental maps revealing spatial distribution of all five elements in which the white boxes show the regions used for elemental quantification.}
  \label{fig:eelsfe}
\end{figure}

\begin{figure}[htbp]
  \centering
  \includegraphics[width=0.82\linewidth]{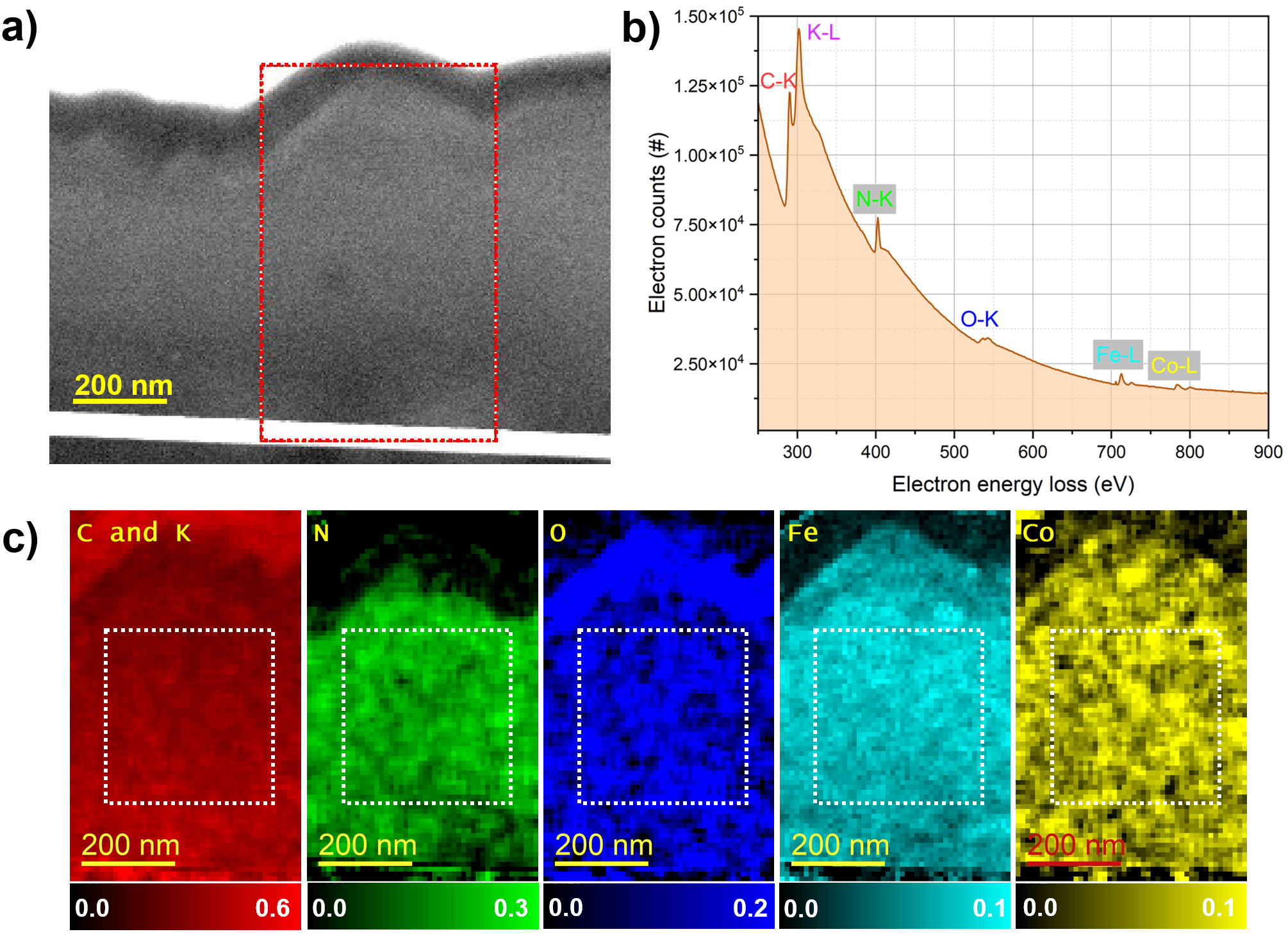}
  \caption{EELS analysis of CoHCF. (a) HAADF-STEM image of the CoHCF lamella with red box indicating the region used for EELS analysis. (b) EEL spectrum showing distinct edges for C, K, N, O, Fe, and Co. (c) Elemental maps revealing spatial distribution of all five elements in which the white boxes show the regions used for elemental quantification.}
  \label{fig:eelsco}
\end{figure}

\begin{figure}[htbp]
  \centering
  \includegraphics[width=0.82\linewidth]{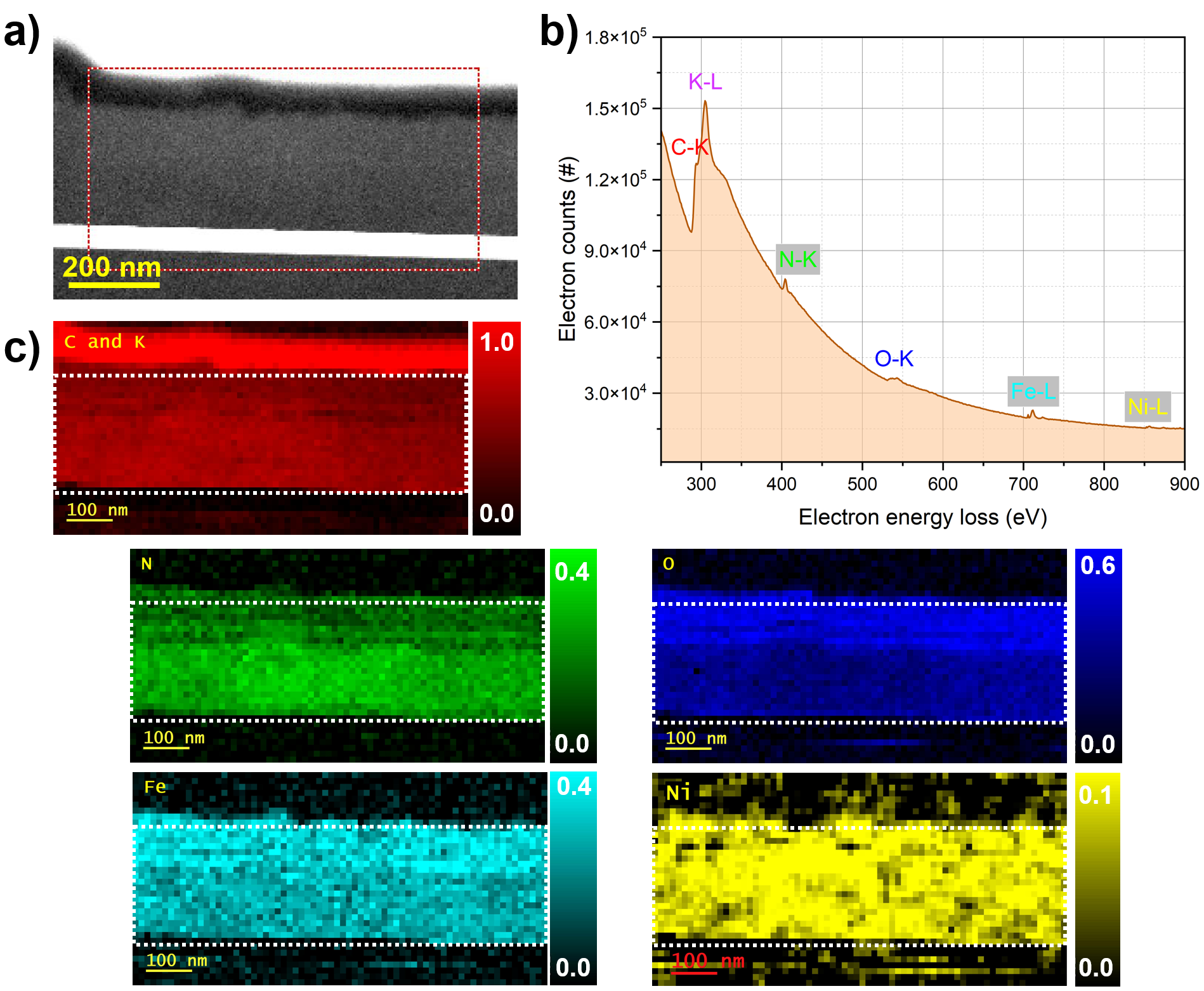}
  \caption{EELS analysis of NiHCF. (a) HAADF-STEM image of the NiHCF lamella with red box indicating the region used for EELS analysis. (b) EEL spectrum showing distinct edges for C, K, N, O, Fe, and Ni. (c) Elemental maps revealing spatial distribution of all five elements in which the white boxes show the regions used for elemental quantification.}
  \label{fig:eelsni}
\end{figure}

These multi-scale TEM observations confirm that all films have crystalline cubic frameworks. While FeHCF and CoHCF exhibits excellent stability under the electron beam, structural disturbances in the NiHCF film are observed, consistent with the SEM investigations presented earlier. To complement these structural insights, elemental quantification and mapping were performed using STEM-EELS analysis. The results are shown in Figures~\ref{fig:eelsfe}, \ref{fig:eelsco}, and \ref{fig:eelsni}.

The HAADF-STEM image (Figure~\ref{fig:eelsfe}a) shows the FeHCF cross-sectional lamella with its surface covered by FIB platinum at the upper right edge. The red-marked region was selected for EELS analysis. The corresponding EEL spectrum exhibits distinct edges for C--K, K--L, N--K, O--K, and Fe--L. However, to generate elemental maps from this spectrum, the significant overlap between the C--K and K--L edges presents a limitation as these edges are closely spaced. Therefore, for all three films analyzed, the spatial distribution of carbon and potassium are mapped together. The elemental maps reveal a uniform spatial distribution of all detected elements within the analyzed area. A slightly higher oxygen concentration is observed near the film surface, which can be attributed to mild surface oxidation upon air exposure.

The HAADF-STEM image (Figure~\ref{fig:eelsco}a) shows the CoHCF cross-sectional sample with the surface oriented at the top. The red rectangle marks the region selected for EELS acquisition. The corresponding EEL spectrum exhibits distinct edges for the key elements: C, K, N, O, Fe, and Co. In the elemental maps, also for CoHCF an oxide enrichment at the surface, accompanied by a depletion of iron and a significant reduction in nitrogen concentration is observed. Underneath this oxidized surface layer of about 50~nm, a virtually uniform spatial distribution of all six elements within the analyzed region is obtained. This uniformity, together with a quantified Fe:Co atomic ratio of about 1:1 (7 at.\% Fe and 7 at.\% Co), confirms the chemical homogeneity and near-stoichiometric composition of the film, consistent with previous EDX measurements.

For NiHCF, the HAADF-STEM image shows the region selected for EELS acquisition, and the corresponding spectrum displays distinct edges for all relevant elements. The elemental maps reveal spatially resolved distributions of C and K (combined), N, O, Fe, as well as Ni, again with an oxidation layer at the surface. The quantified Fe:Ni ratio of approximately 4:3 (4 at.\% Fe and 3 at.\% Ni) is consistent with the expected 1:1 stoichiometry and aligns with previous EDX measurements. Table~\ref{tab:eels} summarizes the elemental quantification obtained by STEM-EELS analysis. It should be noted, however, that these results are affected by the critical need for background subtraction, which may introduce variation up to one percent.

\begin{table}[htbp]
\centering
\caption{Elemental quantification (atomic \%) of FeHCF, CoHCF and NiHCF thin films obtained by STEM-EELS analysis.}
\label{tab:eels}
\begin{tabular}{lrrrrrrr}
\toprule
Material & C (\%) & N (\%) & O (\%) & Fe (\%) & Co (\%) & Ni (\%) & K (\%)\\
\midrule
FeHCF & 42.4 & 28.8 & 8.8 & 12.3 & --- & --- & 14.9\\
CoHCF & 33.6 & 18.7 & 15.4 & 7.3 & 6.8 & --- & 18.3\\
NiHCF & 39.1 & 21.7 & 21.3 & 4.2 & --- & 3.3 & 10.5\\
\bottomrule
\end{tabular}
\end{table}

\begin{figure}[htbp]
  \centering
  \includegraphics[width=0.90\linewidth]{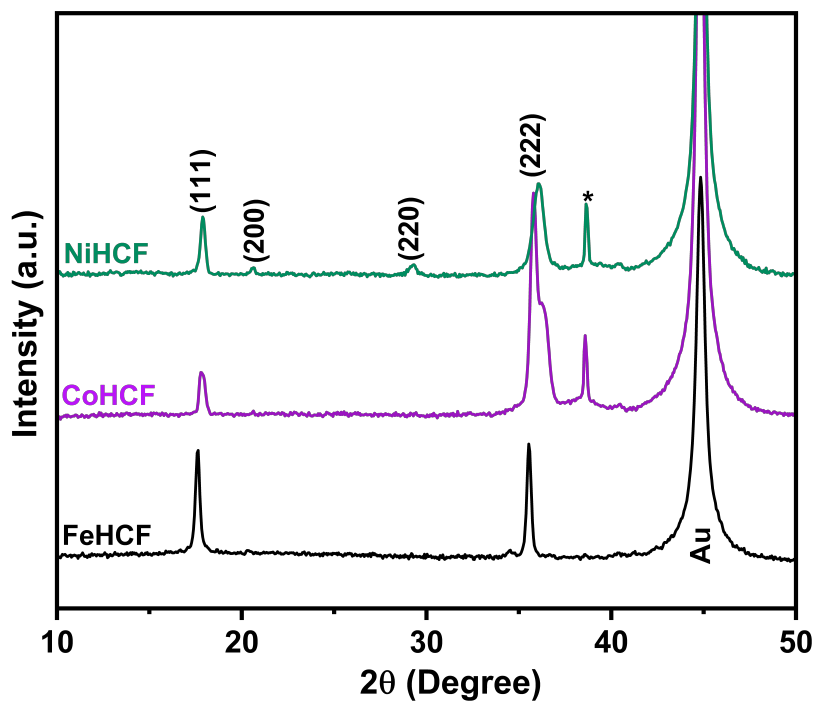}
  \caption{X-ray diffraction pattern of films grown at 0.3 V by potentiostatic deposition.}
  \label{fig:xrd}
\end{figure}

To provide a structural overview, the XRD patterns of FeHCF, CoHCF, and NiHCF are presented in Figure~\ref{fig:xrd}. For FeHCF, characteristic reflections were observed at 17.61$^\circ$ and 35.56$^\circ$, which can be indexed to the (111) and (222) planes of the cubic phase. CoHCF exhibits similar reflections at 17.77$^\circ$ and 35.80$^\circ$, also corresponding to the (111) and (222) planes. In the case of NiHCF, additional diffraction peaks appeared at 20.56$^\circ$ and 29.28$^\circ$, together with the reflections at 17.85$^\circ$ and 36.10$^\circ$, which were indexed to the (111), (200), (220), and (222) planes of the cubic phase. Calculated lattice parameters revealed a gradual contraction from 10.13~\AA (FeHCF) to 10.06~\AA (CoHCF) to 10.01~\AA (NiHCF). This trend is consistent with the decreasing ionic radius from Fe to Ni. The pronounced peak broadening, alongside that attributable to grain size observed in NiHCF and the potential split signal for Co at 35.7$^\circ$ indicate increased microstrain and the presence of point defects \cite{r47,r48,r49}.

A weak additional reflection at approximately 38.0$^\circ$ (marked by an asterisk) is observed exclusively in CoHCF and NiHCF samples and may originate from the formation of oxides and hydroxides (e.g. $\alpha$-Fe$_2$O$_3$/Co(OH)$_2$/Ni(OH)$_2$) upon air exposure \cite{r50}. We attribute this observation to surface oxidation of the films upon exposure to ambient atmosphere that we observed also via STEM-EELS based elemental mapping, as reported previously. A similar mechanism was described in literature for Prussian white \cite{r51,r52,r53}. Importantly, no clear signatures of such species are detected in Raman or XPS, suggesting that the oxidized phase is just distributed at the surface. This interpretation is consistent with SEM/TEM-EDS observations showing a slight oxygen enrichment near the film surface without alteration of the bulk lattice.

\begin{figure}[htbp]
  \centering
  \includegraphics[width=0.82\linewidth]{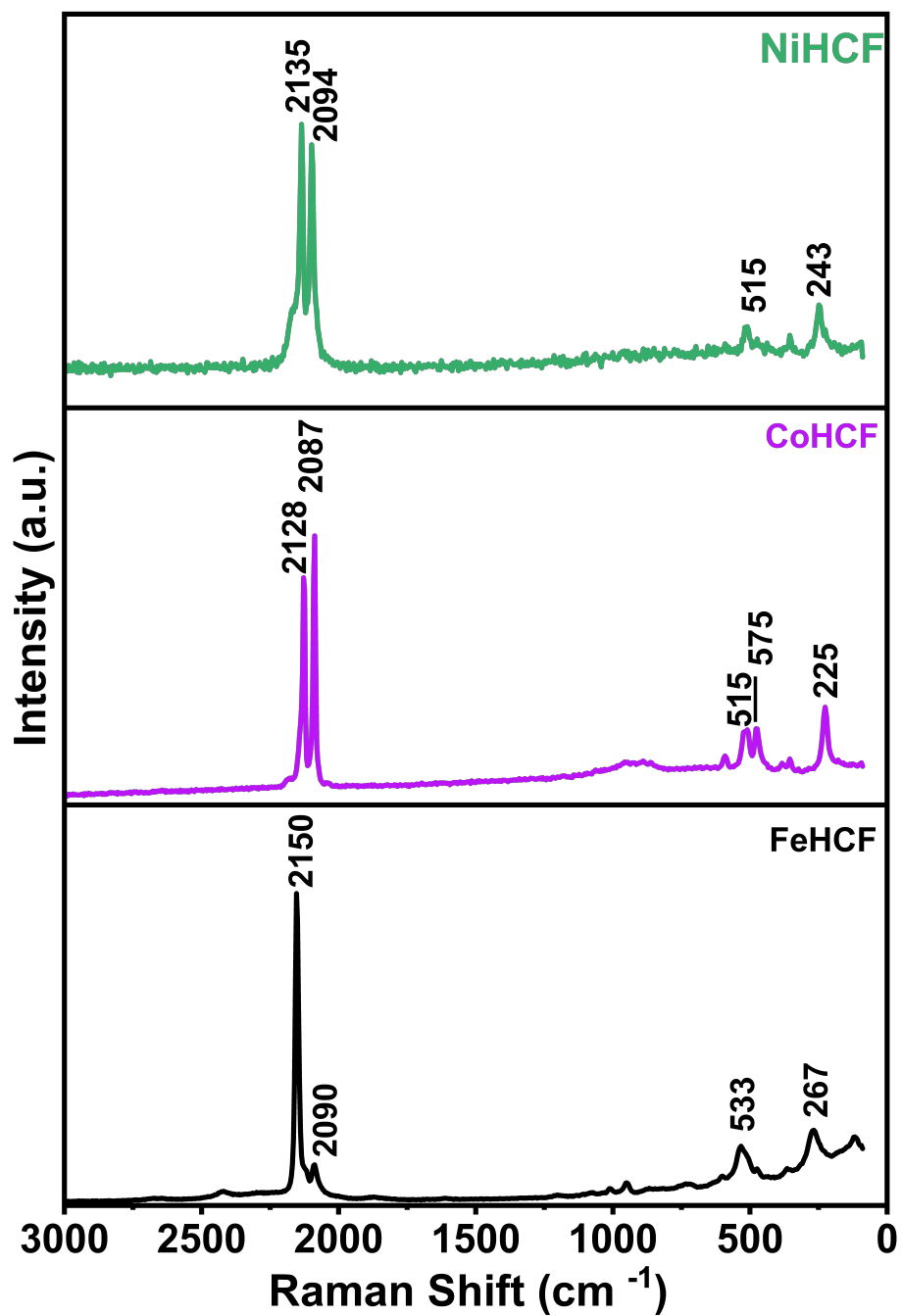}
  \caption{Raman spectra for FeHCF, CoHCF and NiHCF thin films electrodeposited on Si/Au substrate.}
  \label{fig:raman}
\end{figure}

To further investigate the structural framework and possible oxidation effects Raman spectroscopy was performed on FeHCF, CoHCF, and NiHCF thin films. Figure~\ref{fig:raman} shows the Raman spectra of HCF samples, which exhibit the characteristic C$\equiv$N stretching vibrations of Prussian blue in the range of 2070--2200~cm$^{-1}$. Incorporation of cobalt or nickel into the structure resulted in a shift of the C$\equiv$N peak to lower wavenumbers. These shifts suggest a lengthening or weakening of the C$\equiv$N bonds due to changes in the local electron density, influenced by variations in the oxidation state of iron and the electronegativity of the substituting metals, which affect bond lengths and charge transfer inside the cubic lattice \cite{r47,r54,r55}. NiHCF also exhibited broader and less intense Raman signals, indicative of reduced crystallinity, atomic disorder, and the presence of cracks that may cause substrate interference \cite{r56}. Lower-wavenumber bands, observed between 190 and 620~cm$^{-1}$, correspond to metal--CN--Fe bond deformation (190--340~cm$^{-1}$), metal--C stretching vibrations (450--620~cm$^{-1}$), and metal--C vibrations around 400~cm$^{-1}$; metal--O vibrations are also typically found within this region. However, no significant signals were detected, supporting the hypothesis that the oxide formation is restricted to a superficial/protective layer. Shifts and changes in these bands reflect structural distortions and heterogeneity in the local bonding environment caused by metal substitution and lattice strain. Overall, the comparison with pristine Prussian blue indicates that Co and Ni incorporation modifies both the electronic structure and the local geometry of the HCF framework \cite{r55,r57,r58}.

\begin{figure}[htbp]
  \centering
  \includegraphics[width=0.82\linewidth]{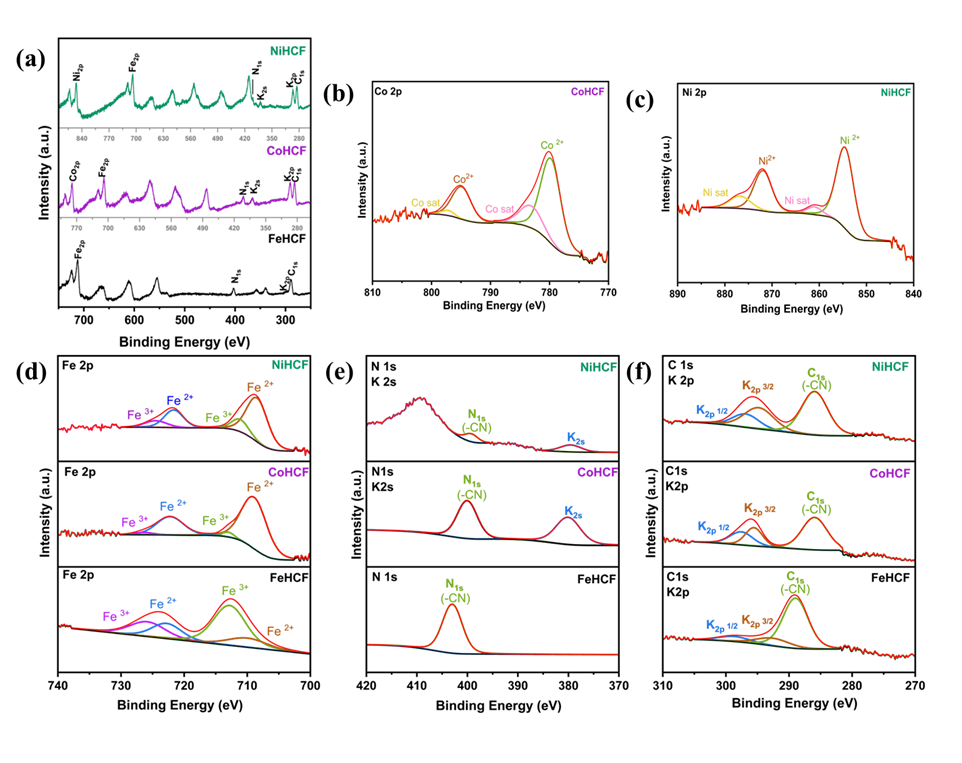}
  \caption{(a) Wide scan spectrum of FeHCF, CoHCF and NiHCF; high-resolution spectra of (b) Co 2p, (c) Ni 2p, (d) Fe 2p, (e) C 1s and K 2p, and (f) K 2s and N 1s.}
  \label{fig:xps}
\end{figure}

X-ray photoelectron spectroscopy (XPS) measurements were conducted to analyze the valence states of Fe, Co, and Ni in hexacyanoferrate (HCF) samples. A comparison of XPS wide scan spectrum of FeHCF, CoHCF and NiHCF is given in Figure~\ref{fig:xps}. High-resolution XPS spectra of C 1s, K 2p, K 2s, N 1s, Fe 2p, Co 2p and Ni 2p are shown in Figure~\ref{fig:xps}. To further analyze the origins of the acquired spectra, peak deconvolution was carried out to investigate metal valence states in the prepared HCF layers. High-resolution Co 2p and Ni 2p spectra confirmed the presence of Co$^{2+}$ and Ni$^{2+}$ via characteristic spin-orbital splitting and satellite peaks. Figure~\ref{fig:xps} shows the existence of Co$^{2+}$ at 780.0 and 795.0~eV, assigned to spin orbitals of Co$^{2+}$ 2p$_{3/2}$ and Co$^{2+}$ 2p$_{1/2}$, respectively. Similarly, Ni 2p shows peaks at 854.6 and 872~eV assigned to Ni$^{2+}$ 2p$_{3/2}$ and 2p$_{1/2}$ spin orbitals. The presence of satellite peaks from Ni 2p (861.0 and 876.8~eV) and Co 2p (783.5 and 797.4~eV) can be explained by charge transfer from the CN ligand to the metal \cite{r59,r60,r61}.

The characteristic mixture of Fe$^{3+}$ and Fe$^{2+}$ is present in FeHCF (Figure~\ref{fig:xps}) with peak maxima at 712.8~eV 2p$_{3/2}$/726.0~eV 2p$_{1/2}$ for Fe$^{3+}$ and 710.3~eV 2p$_{3/2}$/722.9~eV 2p$_{1/2}$ for Fe$^{2+}$. Otherwise CoHCF and NiHCF samples are composed mostly of Fe$^{2+}$ \cite{r62,r63}. The presence and increase of potassium K 2s in the XPS spectra of Co and NiHCF is attributed to the interstitial occupancy of K$^+$ species in the crystal structure \cite{r64}; these results are in agreement with the previous characterization techniques suggesting that the PBAs are predominantly in the reduced Prussian White state. This reduced state requires a charge neutrality within the framework, which is balanced by an increase of K$^+$ and water species observed in elemental characterization as consequence of the reduced Fe$^{2+}$-rich lattice \cite{r65}. Recent nanoscale work on electrodeposited PB analogues further supports the importance of K$^+$ redistribution for electrical conductance modulation \cite{r67}.

\begin{table}[htbp]
\centering
\caption{XPS binding energy assignments for FeHCF, CoHCF, and NiHCF films in eV.}
\label{tab:xps}
\begin{tabular}{lrrr}
\toprule
Peak & FeHCF & CoHCF & NiHCF\\
\midrule
Fe$^{2+}$ 2p$_{3/2}$ & 710.3 & 709.3 & 708.7\\
Fe$^{2+}$ 2p$_{1/2}$ & 722.9 & 722.3 & 721.8\\
Fe$^{3+}$ 2p$_{3/2}$ & 712.8 & 713.2 & 711.5\\
Fe$^{3+}$ 2p$_{1/2}$ & 726.0 & 726.5 & 724.7\\
Co$^{2+}$ 2p$_{3/2}$ & --- & 779.9 & ---\\
Co$^{2+}$ 2p$_{1/2}$ & --- & 795.0 & ---\\
Ni$^{2+}$ 2p$_{3/2}$ & --- & --- & 854.6\\
Ni$^{2+}$ 2p$_{1/2}$ & --- & --- & 872.0\\
K$^+$ 2p$_{3/2}$ & 293.2 & 295.6 & 295.0\\
K$^+$ 2p$_{1/2}$ & 299.1 & 297.6 & 297.3\\
K 2s & --- & 380.0 & 379.5\\
\bottomrule
\end{tabular}
\end{table}

\section{Conclusion}
Thin films of iron, cobalt and nickel hexacyanoferrates were successfully electrodeposited under potentiostatic conditions. Structural and compositional analyses revealed that FeHCF and CoHCF formed films with pyramidal crystal morphology at the surface, whereas NiHCF exhibited less-faceted clustered films with cubic morphology at the surface. Nevertheless, all samples displayed a homogeneous elemental distribution under a ca. 50~nm thick oxidized layer. XRD confirmed a cubic lattice framework across all samples, with progressive lattice contraction observed from Fe to Co to Ni, attributed to ionic radius differences. CoHCF and NiHCF showed the formation of a metal oxide layer on top of the films due to exposure to ambient conditions. Raman and XPS analyses further indicated a shift toward Fe$^{2+}$ dominance and changes in coordination environment in Co- and Ni-substituted films associated with increased interstitial K$^+$ that maintain the charge neutrality and significant presence of water. These findings emphasize the tunability of structural and electrochemical properties via metal substitution in PBAs. Future work will focus on correlating these differences with electrical performance metrics, in line with recent studies of electrodeposited PB-based memristive systems \cite{r23,r27,r66,r67}.

\section*{Author Contributions}
Writing---original draft preparation, L.O.G., M.P. and M.K.F.; writing---review and editing, L.O.G., M.P., M.K.F., D.W., F.R. and C.K.M.; validation, L.O.G., M.P., M.F.K. and D.W.; investigation, L.O.G., M.P., M.K.F. and F.R.; supervision, C.K.M., D.W. and A.L. All authors have read and agreed to the published version of the manuscript.

\section*{Funding}
This research was funded by Deutsche Forschungsgemeinschaft (DFG)---Project number 531524052. This research was funded by the European Social Fund (ESF) and the Free State of Saxony within the framework of the Landesinnovationspromotion program, application number 100670500.

\section*{Acknowledgments}
The authors would like to thank Birgit Opitz from IFW Dresden for support with XRD measurements and Dirk Hildebrand from WHZ for assistance with XPS measurements.

\section*{Conflicts of Interest}
The authors declare no conflicts of interest.

\section*{Abbreviations}
PB, Prussian blue; PBA, Prussian blue analogue; HCF, hexacyanoferrate; CV, cyclic voltammetry; EDX, energy-dispersive X-ray; XRD, X-ray diffraction; FIB, focused ion beam; TEM, transmission electron microscopy; EELS, electron energy loss spectroscopy; STEM, scanning transmission electron microscopy; HRTEM, high-resolution transmission electron microscopy; BFTEM, bright-field transmission electron microscopy; HAADF, high-angle annular dark field; FFT, fast Fourier transform; XPS, X-ray photoelectron spectroscopy.

\end{document}